\documentclass[12pt,preprint]{aastex}
\usepackage[usenames]{color}

\shorttitle{Multiple Plasmoid Ejections and Associated Hard X-ray Bursts}
\shortauthors{Nishizuka et al.}

\begin{document}

\title{Multiple Plasmoid Ejections and Associated Hard X-ray Bursts in the 2000 November 24 Flare}


\author{N. Nishizuka\altaffilmark{1}, H. Takasaki\altaffilmark{1,2}, 
A. Asai\altaffilmark{3,4} and K. Shibata\altaffilmark{1}
}

\altaffiltext{1}{Kwasan and Hida observatories, Kyoto University, Yamashina, Kyoto 607-8471, Japan; nisizuka@kwasan.kyoto-u.ac.jp}
\altaffiltext{2}{Accenture Japan, Ltd., Akasaka Inter City, Akasaka, Minato-ku, Tokyo 107-8672, Japan}
\altaffiltext{3}{Nobeyama Solar Radio Observatory, National Astronomical Observaotry of Japan, Minamisaku, Nagano 384-1305, Japan}
\altaffiltext{4}{The Graduate University for Advanced Studies (SOKENDAI), Miura, Kanagawa 240-0193, Japan}


\begin{abstract}
The Soft X-ray Telescope (SXT) on board Yohkoh revealed that the ejection of X-ray emitting plasmoid is sometimes observed in 
a solar flare. It was found that the ejected plasmoid is strongly accelerated during a peak in the hard X-ray emission of the 
flare. In this paper we present an examination of the GOES X 2.3 class flare that occurred at 14:51 UT on 2000 November 24. 
In the SXT images we found ``multiple'' plasmoid ejections with velocities in the range of 250-1500 km s$^{-1}$, which showed 
blob-like or loop-like structures. Furthermore, we also found that each plasmoid ejection is associated with an impulsive 
burst of hard X-ray emission. Although some correlation between plasmoid ejection and hard X-ray emission has been discussed 
previously, our observation shows similar behavior for multiple plasmoid ejection such that each plasmoid ejection occurs 
during the strong energy release of the solar flare. As a result of temperature-emission measure analysis of such plasmoids, 
it was revealed that the apparent velocities and kinetic energies of the plasmoid ejections show a correlation with the peak 
intensities in the hard X-ray emissions. 
%
\end{abstract}


%

\section{Introduction}
The Soft X-ray Telescope \citep[SXT;][]{tsu91} on board Yohkoh \citep{oga91} revealed that a soft X-ray emitting plasma 
ejection, or plasmoid ejection, is sometimes observed in solar flares \citep[e.g.][]{shi95}. It was also found that the 
plasmoids show blob-like or loop-like shapes and that the strong acceleration of the plasmoid ejection occurs during the 
peak time of the hard X-ray emission \citep{tsu97, ohy97}. Their ejection velocities are typically several hundred km 
s$^{-1}$ and the ejected plasma is heated to more than 10 MK before the onset of the ejection \citep{ohy97, ohy98}. They 
often start to rise up gradually a few tens of minutes before the onset of a hard X-ray burst and are then strongly 
accelerated just before or at the impulsive phase of the flare. A similar kinetic evolution is also seen in the case 
of coronal mass ejections \citep[CMEs;][]{zha01, kim05a, kim05b}.

Similarly, slowly drifting radio structures, observed at the beginning of the eruptive solar flares in the 
0.6-1.5 GHz frequency range, have been interpreted as the radio signatures of plasmoid ejection \citep{kli00, kha02, karl02}. 
Hudson et al. (2001), moreover, identified a rapidly moving hard X-ray source associated with a moving microwave source and 
an X-ray plasmoid ejection. Kundu et al. (2001) also identified moving soft X-ray ejecta associated with moving decimetric/metric 
radio sources observed by the Nan\c{c}ay radioheliograph. Sui et al. (2004) also found a plasmoid ejection 
in hard X-ray images with RHESSI satellite.

In the standard model of solar flares, so-called  CSHKP model \citep{car64, stu66, hir74, kop76}, a filament/plasmoid 
ejection is included. However, it does not necessarily stress the importance of the role of plasmoid ejection explicitly. 
Shibata et al. (1995) and Shibata \& Tanuma (2001) extended the CSHKP model by unifying reconnection and plasmoid 
ejection and stressed the importance of the plasmoid ejection in a reconnection process. The model is called the 
``plasmoid-induced-reconnection'' model. In that model, the plasmoid inhibits reconnection and stores magnetic energy in a 
current sheet. Then, once it is ejected, inflow is induced because of the mass conservation, resulting in the enhancement of 
reconnection rate and the acceleration of the plasmoid due to the faster reconnection outflow. Moreover, reconnection theories 
predict several plasmoids of various scales are generated. The dynamics of plasmoid formation in the solar flare and their 
subsequent plasmoid ejection affect the reconnection rate in the nonlinear evolution. Therefore, plasmoid ejections are 
observational evidence of magnetic reconnection of solar flares. Since plasmoid ejections have been observed in both long 
duration events and compact flares \citep{shi95}, it is shown that the magnetic reconnection model may be applicable even for 
the compact flares that do not show the other typical features of the magnetic reconnection.

On the basis of the results of magnetohydrodynamic (MHD) numerical simulations, Kliem et al. (2000) suggested that each 
individual burst in the slowly pulsating structure is generated by suprathermal electrons, accelerated in the peak of the 
electric field in the quasi-periodic and bursting regime of the magnetic field reconnection. This is the 
so-called ``impulsive bursty'' reconnection \citep{pri85}. In that regime, several plasmoids can be formed successively 
as a result of the tearing and coalescence instabilities \citep{fin77, kli00}. The repeated formation of magnetic islands 
can induce magnetic reconnection and their subsequent coalescence \citep{taj87}. These processes even have a cascading form: 
secondary tearing, tertiary tearing, and so on, always on smaller and smaller spatial scales \citep{tan01, shi01}. Furthermore 
the formed plasmoids can merge into larger plasmoids. Tanuma et al. (2001) also showed that an increase in the velocity of the 
plasmoid ejection leads to an increase in the reconnection rate, and B\'{a}rta et al. (2008) analyzed the dynamics of plasmoids 
formed by the current sheet tearing.


The unsteady reconnection mentioned above can release a large amount of energy in a quasi-periodic way. The energy released 
in the upward direction can be observed as plasmoid ejections or coronal mass ejections, while that in the downward direction 
as downflows \citep{mck99, asa04a} and impulsive bursts at the footpoints of the coronal loops. Bursty energy 
release in solar flares has been observed as highly time variable hard X-ray bursts and microwave bursts \citep{fro71, den85, 
kan70, kip83, ben92, asc02}. Benz \& Aschwanden (1992) and Aschwanden (2002) argued that these impulsive bursts suggest the 
existence of highly fragmented particle acceleration regions. This fragmented structure of solar flares indicates that a flare 
is an ensemble of a vast amount of small-scale energy releases and the fractal/turbulent structure of the current sheet can be 
expected \citep[see also][]{nis09}. Recently the kinematics of multiple plasmoids have been studied by full-particle simulations 
and how the particles interact with their surroundings has been explained \citep[e.g.][]{dra06, pri08, dra09, 
dau09}. It is interesting to note that the stochastic acceleration mechanism \citep[e.g.][]{ben87, bro85, mil96, 
liu06} may be related to particle acceleration in fractal/turbulent current sheet \citep[see also][]{str88, kow09, sam09}.


Karlick\'{y} et al. (2004) showed a unique series of slowly drifting structures during one flare, from which 
the authors proposed that it indirectly maps a formation of several plasmoids and their interactions. However in most of the 
previous studies, only one plasmoid or one drifting structure was reported during the solar flare. In this paper, we present 
for the first time the direct observations of multiple X-ray emitting plasmoid ejections 
associated with a single solar flare observed by Yohkoh/SXT (firstly reported by Takasaki 2006). In section 2, 
we describe the multiple plasmoid ejection events. Then we analyzed the data in section 3 by examining in detail the relationship 
between the multiple plasmoid ejections and the nonthermal hard X-ray emissions using Yohkoh data. Finally we discuss the 
dynamic features of magnetic reconnection and the roll of plasmoid ejections in the particle acceleration in a solar flare 
in section 4.
%
\section{Observation}
\subsection{Overview}
A series of homologous flare-CME events occurred in NOAA Active Region 9236 from 2000 November 24 to November 26. 
These events have been reported by several researchers. Nitta \& Hudson (2001) 
found that the CME-flare events of the homologous flares show quite similar characteristics in both their coronal/photospheric 
magnetic structures and their CME properties. Zhang \& Wang (2002) compared the homologous flares in detail through the use 
of multiwavelength observations. 
Wang et al. (2002) 
reported that the activities of these flares was driven or triggered by newly emerging magnetic flux, which appeared on 
the western side of the leading sunspot in this active region. Figure 1a-1c show snapshot images of the preceding sunspot 
in NOAA 9236 and an associated two-ribbon flare, which occurred on 2000 November 24 observed in white light, ultraviolet 
and soft X-ray emission. Takasaki et al. (2004) performed a comparison of the physical parameters between the individual 
flares and from this they could confirm that the plasmoid-induced-reconnection model is reasonable. They then showed that 
the interaction between the new emerging magnetic flux loops and the pre-existing magnetic field was essential for producing 
the homologous flares and plasmoid ejections in the active region. 

These ejections were followed by a single halo-CME which occurred at 15:30 UT on 2000 November 24. Figure 2 shows a CME 
image observed with the Large Angle Spectroscopic Coronagraph \citep[LASCO;][]{bru95} that occurred following the flare 
\citep[e.g. see][]{nit01, zha02}. The core of the CME was observed traveling in the northwest direction. Moon et al. (2003) 
found a good correlation between CME speed and the GOES X-ray peak flux of the associated flares in this series of homologous 
flare-CME events.

Impulsive hard X-ray bursts were also observed in this flare with the Hard X-ray Telescope \citep[HXT;][]{kos91} on board Yohkoh 
(see Fig. 1d). A pair of hard X-ray sources was located at the footpoints of the coronal arcade. We used the hard X-ray emission 
data observed with the H-band (52.7-92.8 keV) of HXT, whose temporal resolution was 0.5 s.

\begin{figure}
\epsscale{.80}
\plotone{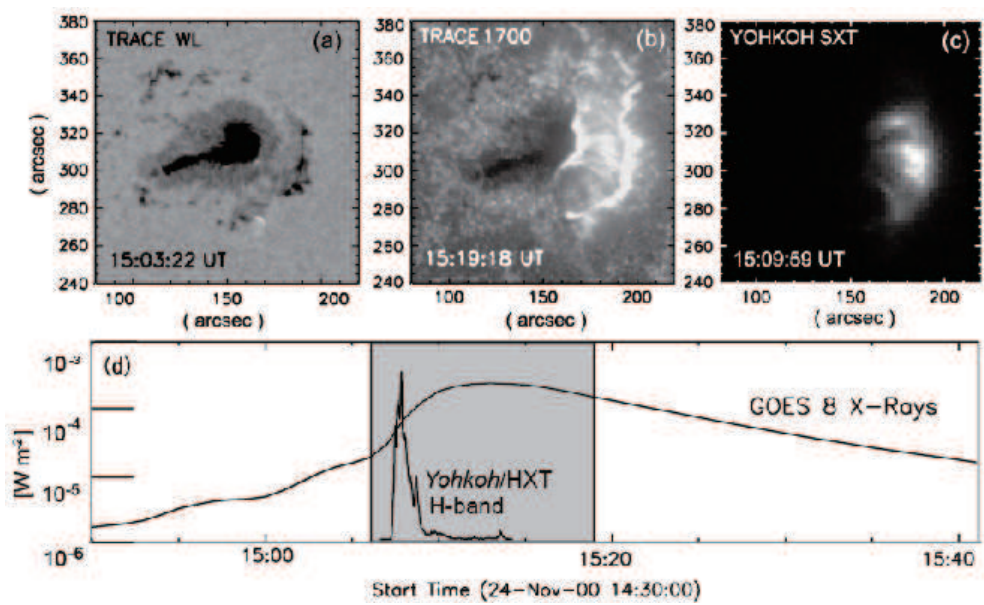}
\caption{(a) Snapshot image of the preceding sunspot of NOAA Active Region 9236 taken with the white light filter of the Transition 
Region and Coronal Explorer \citep[TRACE;][]{han99}. (b) Snapshot image of the two-ribbon flare and the coronal loops located at 
the west side of the sunspot taken with 1700 $\AA$ filter on board TRACE and (c) full-resolution image of Yohkoh/SXT. 
(d) GOES soft X-ray flux and the hard X-ray emission observed with Yohkoh/HXT. \label{fig1}}
\end{figure}

\begin{figure}
\epsscale{.40}
\plotone{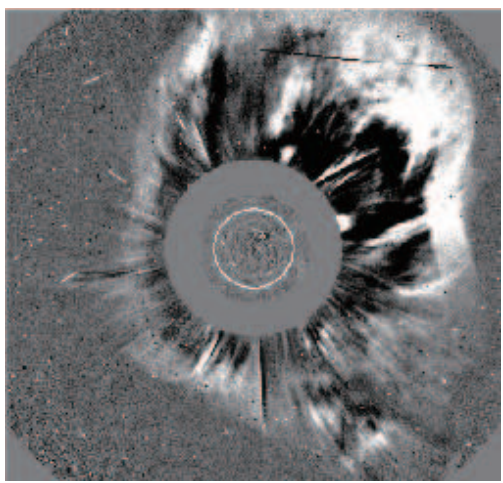}
\caption{Running difference image of the coronal mass ejection (CME) taken with C2 LASCO on board SOHO at 
16:00 UT on 2000 November 24.\label{fig2}}
\end{figure}

\begin{figure}
\epsscale{.40}
\plotone{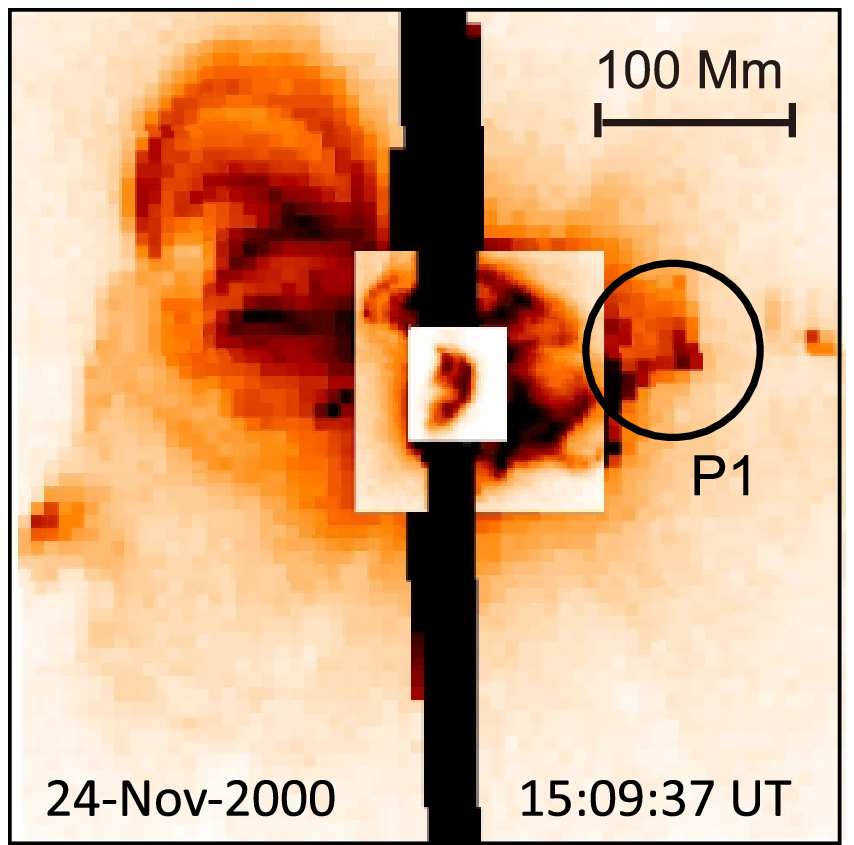}
\caption{Time evolution of the flare which occurred at 14:51 UT on 2000 November 24. This is made of full-, half- 
and quarter-resolution images taken with Yohkoh/SXT AlMg filter (negative images). (Supplement Movie 1; Courtesy 
of J. Kiyohara) \label{fig3}}
\end{figure}

\subsection{Multiple Plasmoid Ejections}
We focus on a GOES X2.3 class flare that occurred in NOAA 9236 (N19$^{\circ}$, W06$^{\circ}$) at 14:51 UT on 2000 November 
24. This flare was one of the homologous flares which were described previously. In this flare, we observed seven plasmoid 
ejections in the soft X-ray images of the flare taken with Yohkoh/SXT. We mainly used the partial frame images with half- 
and quarter-resolution for the analysis. The spatial resolutions are about 5'' and 10'', respectively. We used the sandwich 
(AlMg) filter images which were taken with 20 second cadence. Figure 3 and Supplement Movie 1 show the temporal evolution of 
the flare, which is made of full-, half- and quarter-resolution images taken with Yohkoh/SXT AlMg filter (negative images). 
Full-, half- and quarter-resolution images are different in their spatial resolutions, field of views and 
exposure times. Full-resolution images  are of short exposure time and focus on the brightest region of the active region, 
such as the two ribbon structure. On the other hand, quarter-resolution images are of longer exposure time so that they are 
applicable for the detection of large-scale and faint phenomena such as plasmoid ejections. The black vertical line in the 
middle of Figure 3 shows the saturation of the quarter-resolution images. We identified seven major ejections which we named 
P1-P7. In Figure 4 (Supplement Movie 2), Figure 5 and Figure 6, we marked each plasmoid ejection with a 
circle. Each plasmoid ejection can be seen more clearly in quarter-resolution images of Figure 5, 
while half-resolution images of Figure 6 are convenient to see plasmoids just after ejections. The fields of view of 
quarter- and half-resolution images are also shown in Figure 3.

Figure 7 shows the temporal evolution of soft X-ray emission observed with SXT. A full resolution image is inset on Figure 7a.
Figure 7a-7c shows three of the plasmoid ejections denoted as P1, P4 and P7. In Figure 7d-7f we overlaid contour images 
of the soft X-ray emission, which show the time evolution of the plasmoid ejection. The directions of the ejections are 
indicated by the arrows in the panels. To make clear the traveling of the plasmoids, we also overlaid the contours of the 
SXR images taken at different times (e.g. 15:09:19 UT, 15:09:39 UT and 15:09:59 UT for Fig. 7d) on Figures 7d, 7e, and 7f. 
From these contour images, we can roughly outline the position and the size of the bright cores of the plasmoids.  We can 
also measure the velocity $V$ by taking the time difference of these contour images. Those size and apparent 
velocity of the plasmoids are listed in Table 1. Here we note that these are the ``apparent'' velocities, and the motions in 
the line of sight are ignored. Therefore, the actual velocity of the plasmoid $V/\cos\theta$ will be greater than 
the plane-of-the-sky value, where $1/\cos\theta$ representing the expected deprojection over a reasonable range of angles 
$\theta$ to the plane of the sky. The trajectory of the plasmoid is not necessarily in the straight lines as shown with the 
arrows in Figure 7. In order to take into account those non-straight motions, we measured the apparent velocity of each plasmoid 
by averaging the velocities derived from each time differences. In the following discussions, we consider the apparent velocity 
as the actual velocity.
 
We can see several plasmoids were ejected in the northwest direction, and one plasmoid ejected in the southwest direction 
in Figure 4 (supplement movie 2) and Figure 5, which is marked as P6 in this paper. Each plasmoid has a 
unique velocity, brightness and size. The first ejection was that of P1, which was followed by P2-P4 which were successively 
ejected as a group in the same direction. P1 shows blob-like structure, while P2-P4 seem to be a part of an expanding loop. 
We are not able to clearly define the structure of P5 and P6 due to the faint emission, although we can surely identify P5 
and P6 traveling outward from the active region. P7 is the brightest ejection of blob-like structure. It starts to rise up 
gradually at 15:12 UT and ejected/accelerated upwards at 15:14 UT. 

In the LASCO CME images we can no longer identify the fine structure corresponding to the individual plasmoid 
ejections, though some complicated structures can be observed (see Fig.2). This is probably because the ejected plasmoids merge 
into a single CME. The average velocity of the CME listed in the CME event catalog\footnote{CME event catalog: 
http://cdaw.gsfc.nasa.gov/CME\_list/} is about 1245 km s$^{-1}$. This apparent velocity of the CME is faster than those of the 
plasmoid ejections P1-P7 summarized in Table 1. This observational fact qualitatively suggests that the merged plasmoids are 
continuously accelerated as they are ejected into interplanetary space, as shown by Cheng et al. (2003), although we can hardly 
identify the one-to-one relation between them.

\begin{figure}
\epsscale{.40}
\plotone{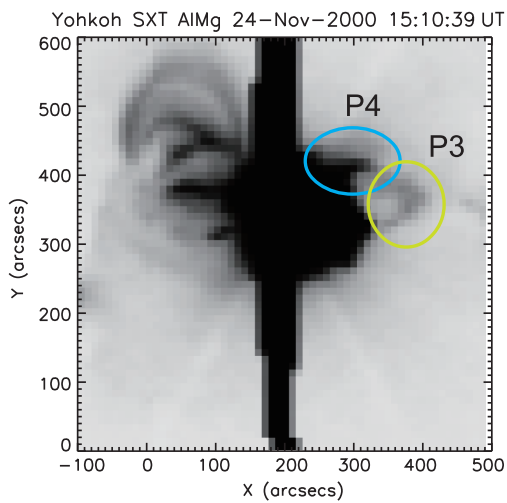}
\caption{The multiple plasmoid ejections associated with the flare on 24 Novermber 2000, which are marked with 
color circles (Supplement Movie 2). This is made of quarter-resolution images taken with Yohkoh/SXT 
AlMg filter (negative images). \label{fig4}}
\end{figure}

\begin{figure}
\epsscale{.80}
\plotone{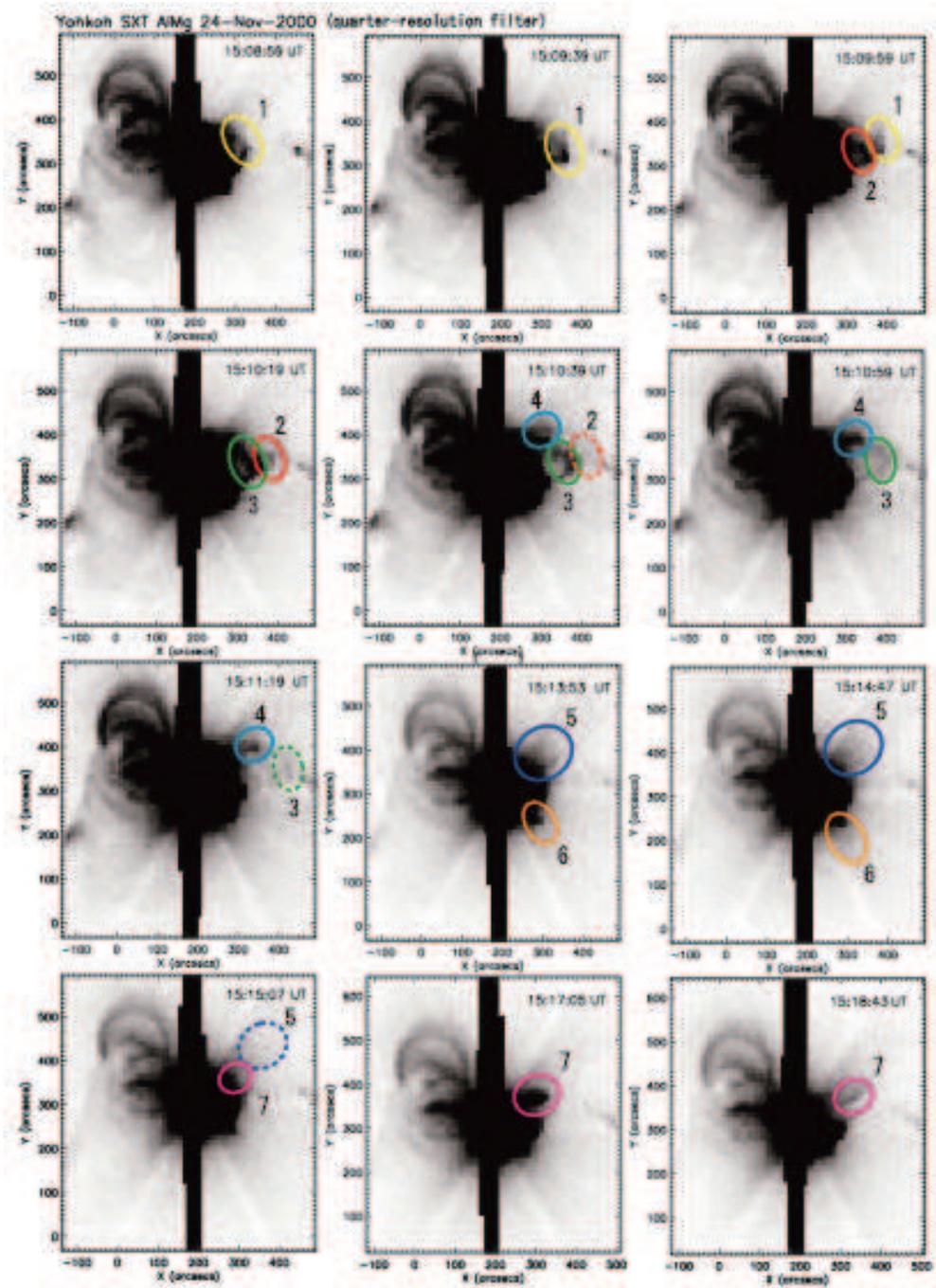}
\caption{The series of snapshot images of multiple plasmoid ejections asociated with a flare on 24 November 
2000. They are quarter-resolutiuon images of the sandwich (AlMg) filters of SXT on board Yohkoh. Each plasmoid ejection is 
marked with a circle and indices of ejection number. \label{fig5}}
\end{figure}

\begin{figure}
\epsscale{.80}
\plotone{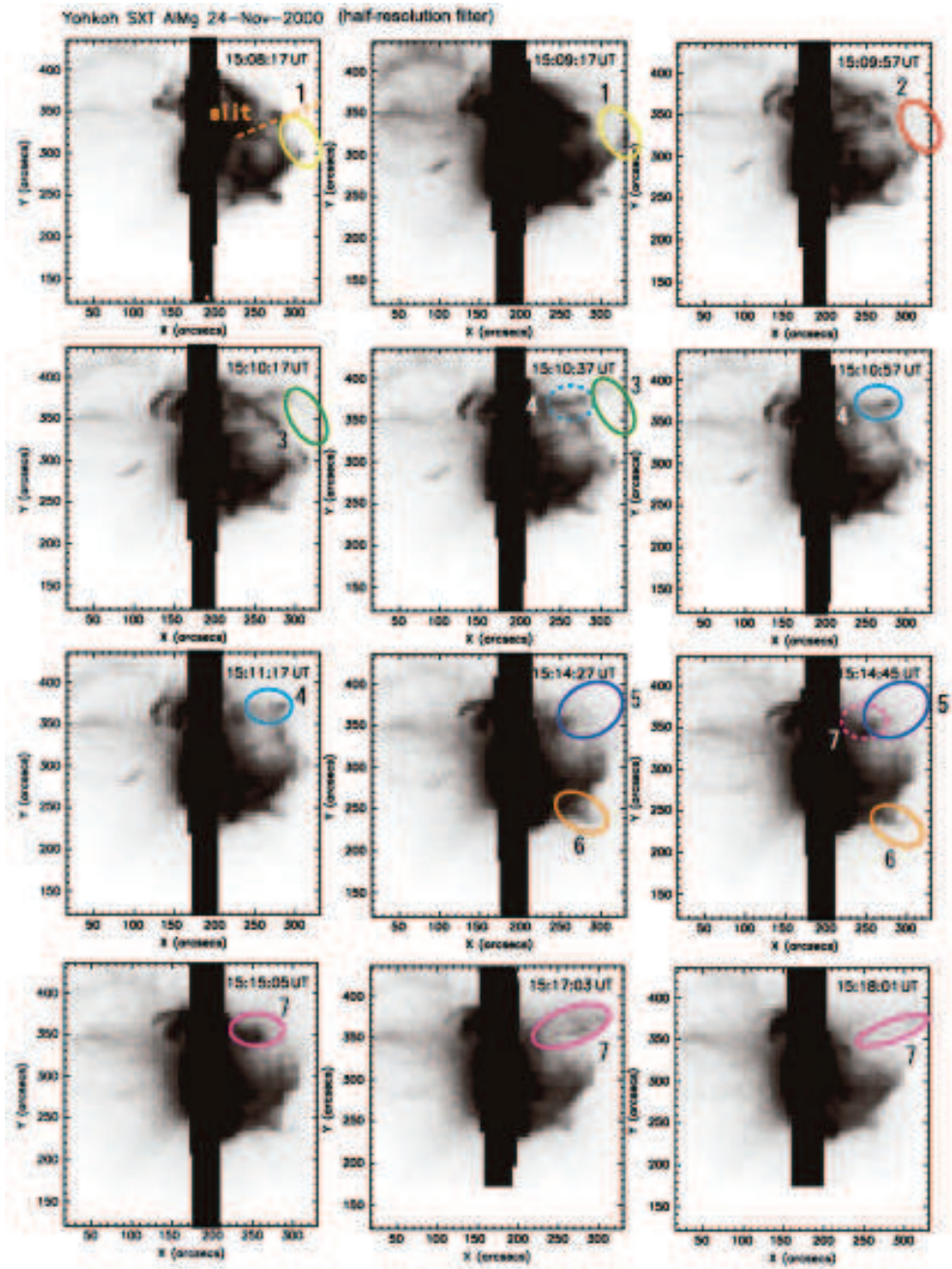}
\caption{The series of snapshot images of multiple plasmoid ejections associated with a flare on 24 November 
2000. They are half-resolution images of the sandwich (AlMg) filters of SXT on board Yohkoh. Each plasmoid ejection is 
marked with a circle and indices of ejection number. Dotted line shows the slit position, along which the time slice image 
of Figure 8 was made. \label{fig6}}
\end{figure}

\begin{figure}
\epsscale{.85}
\plotone{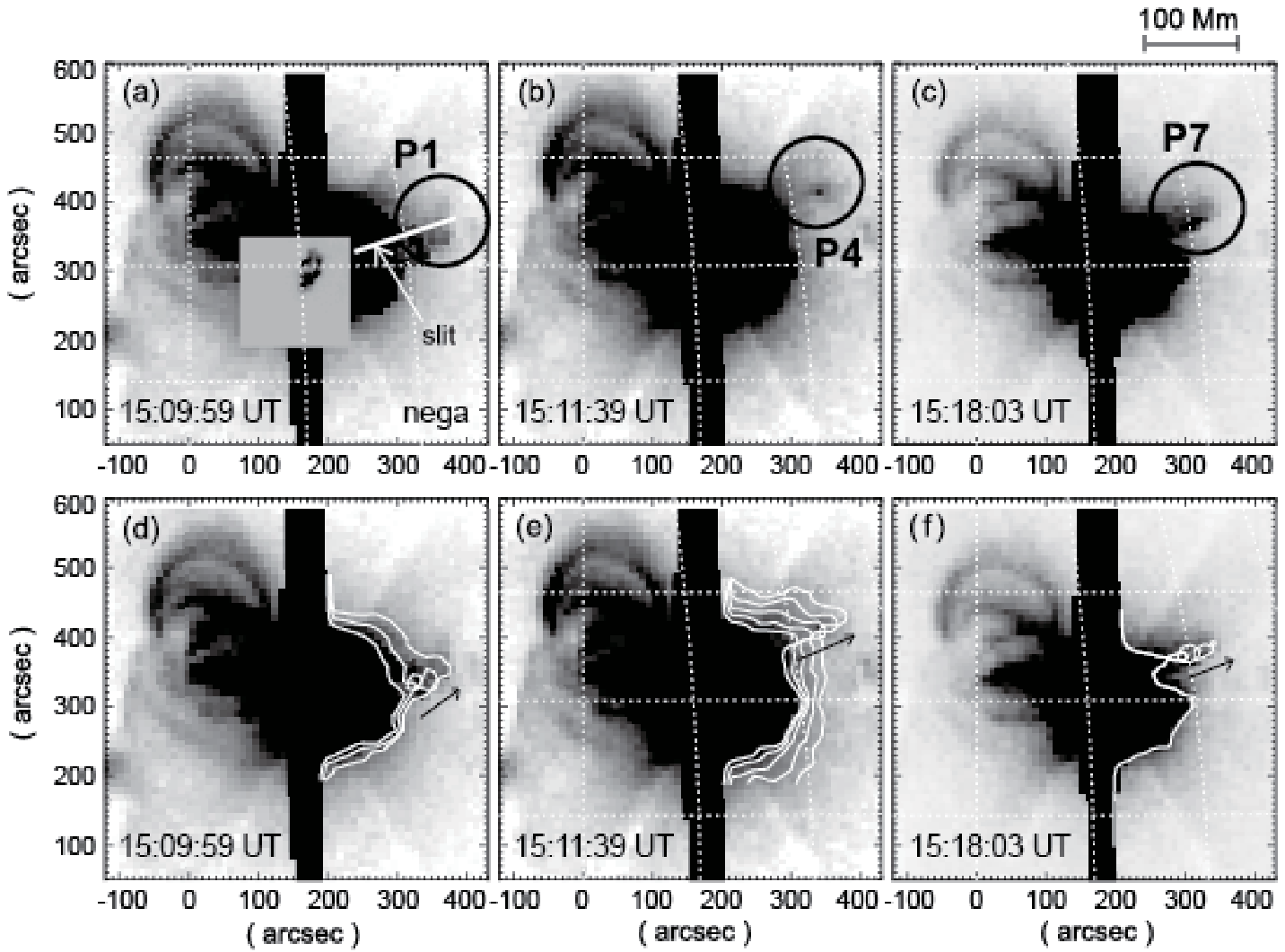}
\caption{Snap shot images of three different plasmoid ejections occurred at 15:00 UT on 2000 November 24 taken with 
Yohkoh/SXT AlMg filter (negative images). (a-c) Quarter-resolution images and a full-resolution image of SXT. Plasmoid 
ejections are denoted with the number of P1, P4 and P7. A full-resolution image in (a) shows the same two-ribbon structure 
as shown in Figure 1. (d)-(f) The time evolution of the plasmoid ejections. Contour plots of the plasmoids of (d) are 
taken at 15:09:19 UT, 15:09:39 UT and 15:09:59 UT, while (e) at 15:10:39 UT, 15:10:59 UT, 15:11:19 UT, 15:11:39 UT 
and 15:11:59 UT, and (f) at 15:16:45 UT, 15:17:23 UT and 15:18:03 UT. \label{fig7}}
\end{figure}
%
%
\section{Analysis and Results}
\subsection{Time Slice Images of Plasmoid Ejections and Comparison to the Hard X-ray Bursts}
Here we focus on the time evolution of the plasmoid ejections. We used a time slice image of the plasmoid ejections as we show 
in Figure 8. The horizontal axis is the time from 15:06 UT to 15:18 UT, and the vertical axis is the 1D image (negative images) 
using a slit line placed along the direction of the several plasmoid ejections (P1, P3, P4, P5 
and P7). This time slice image is made of half-resolution images. The position of the slit line is shown in 
Figure 6 and 7a. Those ejections seen in the time slice image are marked with signs $\square$, 
$\triangle$ $\dots$ in Figure 8b. We can also identify further additional faint ejections in 
the time slice image. Some ejections travel along the slit lines, while others travel on a path which is slightly different 
from the slit line. As a result, the visibility of each plasmoid ejection is different.

Initially P1, P2 and P3 are slowly accelerated, then strongly accelerated during the initial impulsive phase of the hard 
X-ray emission (15:07:40-15:08:40 UT) followed closely by the ejection of P4. A group of plasmoids gradually rise up 15:09:40-
15:12:20 UT followed by the faint ejection P5. P6 is ejected in a different direction (southwest) and does not cross the slit 
line, so it does not appear in Figures 8a,b. P7 is the brightest ejection and the most clearly visible in Figures 8a,b. The 
apparent velocities of the plasmoids along the slit can also be derived from the slopes of the fitted lines in Figure 8b. We 
note that these are the apparent velocities measured from the time slice images of Figure 8b and different from the velocities 
measured from contour plot images in Figure 7. This is because the former shows the front velocity of thinner density plasma, 
while the latter shows the velocity of the thick core part of the plasmoids. In Figure 8a and 8b, the plasmoid ejections 
roughly start to rise at the apparent height of approximately 50'' ($\sim$35 Mm) and propagate into the upper 
atmosphere. This probably means that reconnection occurs at around or just below the height of $\sim$35 Mm. 
In this paper, we set the start position of plasmoid ejections as the height ($\sim$35 Mm) and define the start 
times of ejections as the time when each plasmoid crosses the height (which is shown with dotted line in Fig. 8b). It is noted 
that Figure 6 and Figure 8a,b are drawn with half-resolution images, but Figure 5 is a series of 
quarter-resolution images. Therefore we see more of the earlier phase of plasmoid ejection in Figure 6 and 8 
than in Figure 5, and so Figure 8 is more appropriate to determine the time of plasmoid ejection. We confirmed that the start 
times are comparable to those defined above.

In Figure 8c, we show the light curve of the hard X-ray emission obtained with the H-band (52.7-92.8 keV) of Yohkoh/HXT and 
the GOES soft X-ray light curve. We can distinguish the hard X-ray bursts into three separate periods: the first period, A 
(15:08-15:09 UT), is the brightest phase of the hard X-ray emission, the second period, B (15:09-15:11 UT), 
show gradual enhancement and the last period, C ($\sim$15:14 UT), is an isolated hard X-ray peak. The plasmoid ejections of 
P1-P3 seem to be ejected during the peak time of period A. P1 seems to be ejected just before the hard X-ray 
peaks in Figure 8c. This is consistent with the report by Ohyama \& Shibata (1997) who showed that plasmoids are ejected at or 
just before the hard X-ray peak. During period B, a group of plasmoids gradually rise up and are followed by the faint 
ejection of P4 and P5, while the brightest plasmoid P7 is ejected during period C. The correspondence of the 
plasmoid ejections and hard X-ray peaks are shown by the arrows in Figure 8b. Since the bursts in period A are superposed 
and very complex, it is difficult to identify the exact correspondence between the plasmoid ejections (P1-P3) and the hard 
X-ray bursts.

We made a correlation plot of the times of the hard X-ray bursts and those of the plasmoid ejections (Figure 9a) to make 
the relation clearer. Both the horizontal and vertical axes show the times (UT). The horizontal ({\it light gray}) and the 
vertical ({\it dark gray}) lines illustrate the times of the plasmoid ejections and those of the hard X-ray bursts, 
respectively. The thickness of the lines shows an estimation of the error. Figure 9b shows hard X-ray (52.7-92.8 keV) 
light curves obtained with H-band of Yohkoh/HXT. The thin solid line shows the points where the times of the plasmoid 
ejections correspond to those of the hard X-ray bursts. These results appear to show that several plasmoid ejections 
coincide with hard X-ray bursts.

Here we note that, although the intense HXR emissions indicate that strong energy releases occur at those 
times, it does not necessarily mean that certain amount of plasmoids, that is, notable plasmoids are ejected.  We have 
already noticed that there are many fainter ejections, which also tend to appear at HXR bursts, while some of the hard 
X-ray intensity fluctuations have no associated ejections in Figures 8 and 9. On the other hand, the HXR burst for P5 
and P6 does not show a sharp summit but a gentle hump. This is probably because ejections P5 and P6 are parts of continuous 
outflows from the active region during this time range. This may also suggest a milder energy release compared with the 
others, resulting into a gradual enhancement of hard X-ray emission between 15:09:40 UT and 15:11:00 UT.

\subsection{Temperature Diagnostics of Seven Plasmoid Ejections}

We also studied temperature diagnostics on the soft X-ray emitting plasma using the SXT filter ratio method \citep{har92}. 
We used the half-resolution images taken with the Be and thick Al filters of SXT for the analysis of plasmoid ejections. 
We subtracted the background photon flux of plasmoids. The temperature and emission measure are determined 
using the two filter data of the Be and thick Al filters. The size of the plasmoid $S$ was measured from the contour plot 
images in Figure 7b, which shows the lower limit of the observable size (summarized in Table 1). We assume that the X-ray 
emitting plasma, measured by the filter ratio method, fills the plasmoid with a filling factor of 1. We assumed that the 
volume of the plasmoid is $S^{3/2}$, such that the line of sight width of a plasmoid is equal to the square-root of the 
size $S$. Since the observation times of the Be and thick Al filters are not exactly the same, we used two images from the 
thick Al filter which were taken just before and just after the Be filter observation. The error shown in Table 1 mostly 
results from combining these two images. This error is much greater than the background photon noise \citep[see][in more 
detail]{ohy97}. As for plasmoid ejections P2 and P3, there are no Be and thick Al filter images, because P2 and P3 are out 
of field of view of half-resolution images as shown in Figure 6. In these cases we assumed the temperatures 
of P2 and P3 as $10^7$ K.

Table 1 summarizes the physical parameters, such as temperature, emission measure, density, mass, thermal 
energy and kinetic energy of each plasmoid, of the seven plasmoid ejections identified in Figure 5. Since the size of the 
plasmoids that we derived from the images is just the lower limit, the mass, thermal and kinetic energies calculated with 
the size are also the lower limit. In Table 1, each plasmoid shows a typical temperature of 10$^7$ K, a density of 10$^9$ 
cm$^{-3}$ and an apparent velocity of 200-1400 km s$^{-1}$, which is similar to results of previous studies 
\citep[i.e.][]{ohy97, ohy98}. The kinetic energy of each plasmoid ejection seems to be comparable to or twice as large as 
their thermal energy. We also estimate the total flare energy from the full-resolution images (spatial resolution 2.5") of 
the same filters at the peak time of GOES soft X-ray emission. Since the only part of the total flare energy is converted 
into plasmoid ejections, the total energy of plasmoid ejections is smaller than the total energy of solar flare.

\begin{figure}
\epsscale{.75}
\plotone{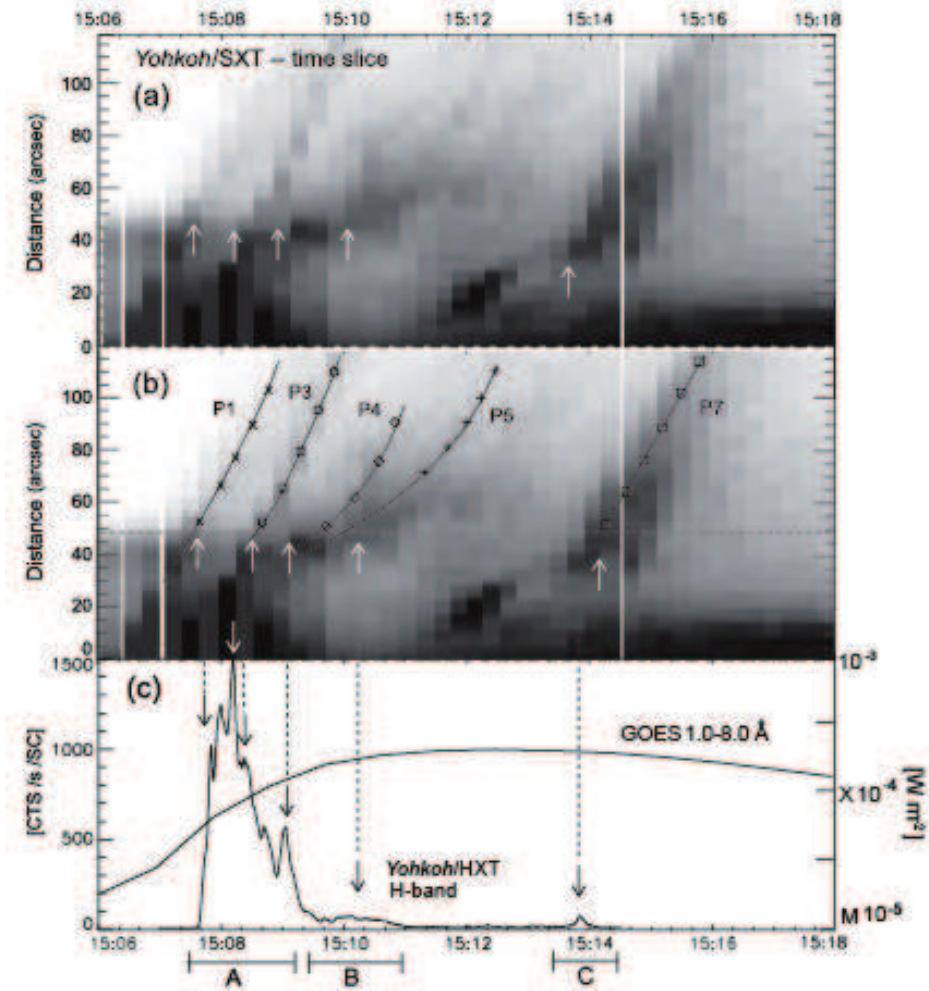}
\caption{(a) Apparent height of the plasmoids ejected in the 2000 November 24 flare at 15:00 UT. Time slice image 
is obtained with half-resolution images of Yohkoh/SXT. The signs of plus (+), circle ($\circ$), triangle 
($\triangle$) and square ($\square$) are plotted over the time slice image in figure(b). (c) Counting rates of hard X-ray 
emission observed with the H-band (52.7-92.8 keV channel) of Yohkoh/HXT and the corresponding GOES soft X-ray light 
curve. \label{fig8}}
\end{figure}

\begin{figure}
\epsscale{.80}
\plotone{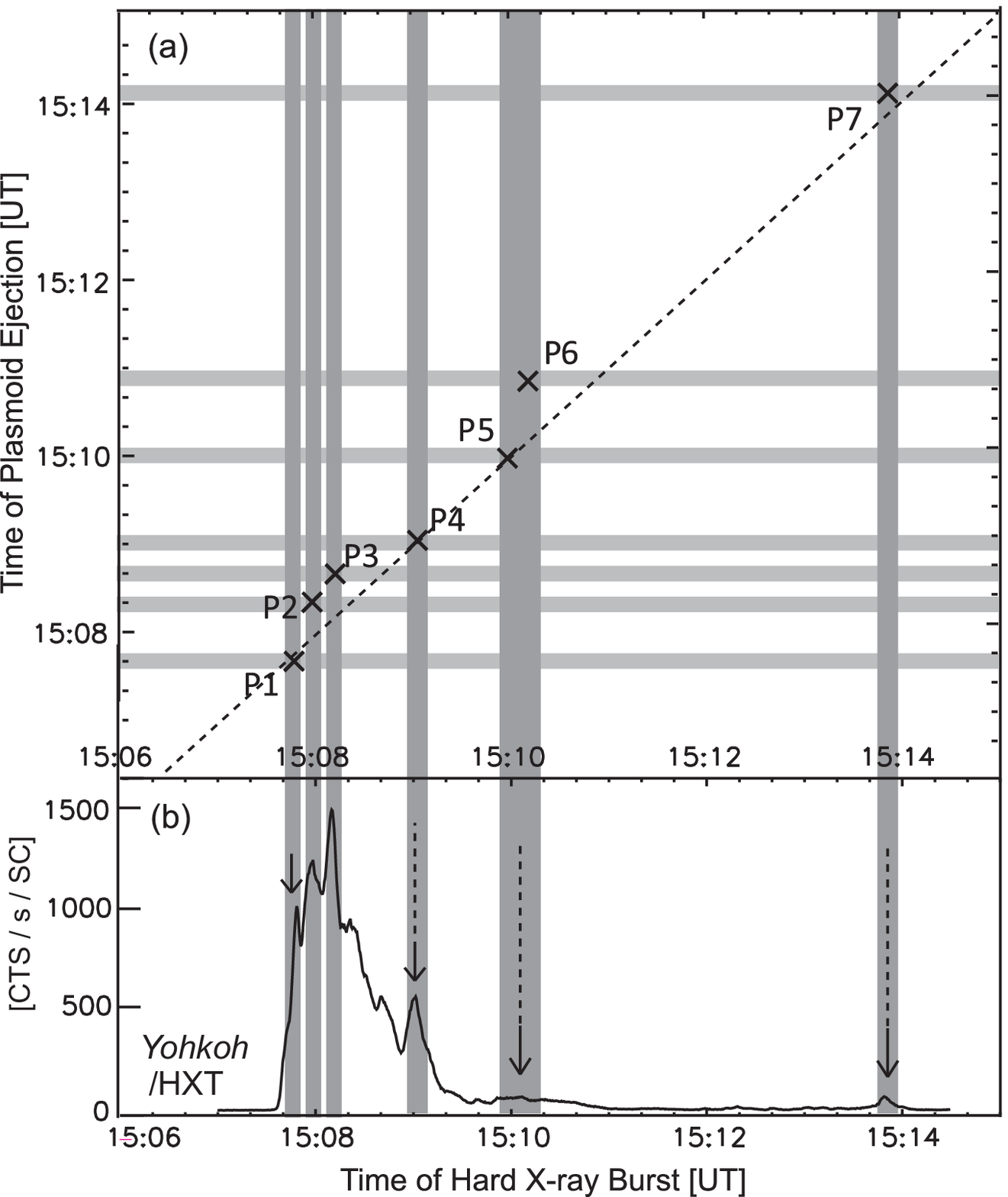}
\caption{(a) Correlation plot between the times of the hard X-ray (HXR) bursts and those of the plasmoid ejections. 
Both the horizontal and vertical axes show the times (UT). The horizontal ({\it light gray}) and the vertical ({\it dark 
gray}) thick lines illustrate the times of the plasmoid ejections and those of the hard X-ray bursts, respectively. The 
width of the thick lines shows the amount of error approximately. (b): HXR (52.7-92.8 keV) light curves 
obtained with H-band of Yohkoh/HXT. The thin solid line shows the points where the times of the plasmoid ejections 
correspond to those of the HXR bursts. \label{fig9}}
\end{figure}

%
%
\section{Summary and Discussion}

We analyzed the GOES X2.3 class flare which occurred at 14:51 UT on 2000 November 24. We found multiple plasmoid ejections 
from a single flare. Furthermore, each plasmoid ejection seems to be associated with a peak in the hard X-ray emission. 

In Figure 8, the plasmoid ejections seem to occur at the height of approximately 50" ($\sim$ 35 Mm). This tells 
us that reconnection occurs at the height of approximately 35 Mm. The horizontal light gray lines in the top 
panel of Figure 9 show that the times of plasmoid ejections, when they reach the height of 35 Mm, seems to be 
well correlated to the peak time of hard X-ray emission, which is consistent with previous studies \citep[e.g.][]{ohy97}. Since 
the peak in hard X-ray emission indicates strong energy release, we have demonstrated that each plasmoid ejection occurs during 
a period of strong energy release, suggesting a series of impulsive energy releases in a single flare.

We also performed a temperature and emission measure analysis and investigated the physical parameters of the plasmoid ejections 
shown in Table 1. Figure 8 and Table 1 show that the hard X-ray bursts in period A have large intensities in correlation with the 
large kinetic energy of plasmoid ejections P1-P4. Conversely, the hard X-ray bursts in period B and C have small intensities in 
correlation with the kinetic energy of plasmoid ejections P5-P7. Figure 10a shows the relation between the kinetic energy of 
plasmoid ejection and the intensity of the corresponding hard X-ray peak emission. We can see a rough tendency that the larger 
kinetic energy of plasmoid ejections is associated with the brighter hard X-ray peak emission, and vice versa. The hard X-ray 
emission is known to show energy release rate \citep[e.g. Neupert effect;][]{neu68, asa04b}, which leads us to the following 
equation:
\begin{equation}
I_{HXR} \Delta t \sim \frac{dI_{SXR}}{dt} \Delta t \propto \frac{dE_{th}}{dt} \Delta t \sim \Delta E_{th} \sim \Delta E_{kin} 
\sim \frac{1}{2}mV_{pl}^2
\end{equation}
where we assumed that the released thermal energy of a solar flare in a short period is comparable to the kinetic energy of 
the plasmoid ejection. This means that a plasmoid ejection with large hard X-ray emission, and therefore with large energy 
release, can be accelerated strongly. A similar kinetic evolution is also seen in the case of CMEs \citep{yas09}.

Figure 10b shows the relation between the apparent velocity and the intensity of the corresponding hard X-ray 
peak emissions. There seems to be a correlation between the plasmoid velocities and the hard X-ray emissions, although it is 
difficult to measure the velocities of the plasmoids precisely due to the faint emission. Similar to the above equations, we 
can derive the following relation between hard X-ray emission and the kinetic energy of a plasmoid ejection: 
\begin{equation}
I_{HXR} \sim \frac{dI_{SXR}}{dt} \propto \frac{dE_{th}}{dt} \sim \frac{B^2}{4\pi}v_{in}L^2 \propto V_{CME} \sim V_{pl} 
\end{equation}
Here $B$ is the typical magnetic field in a current sheet, $v_{in}$ is the inflow velocity and $L$ is the characteristic 
length of the inflow region. The inflow velocity $v_{in}$ is thought to be about 0.01$v_{A}$ from direct observations 
\citep{yok01,nar06}. It has been also known that inflow can be controlled by the plasmoid ejection. As a plasmoid is 
ejected out of the current sheet, the density in the current sheet decreases, and the inflow is enhanced to conserve the 
total mass under the condition that incompressibility is approximately satisfied. Here, we assume that $v_{in}$ is proportional 
to the plasmoid velocity $V_{pl}$. Then, we find the relation $I_{HXR}$ $\propto$ $V_{pl}$. Moreover, if we can further assume 
the $V_{pl}$ is proportional to CME velocity $V_{CME}$, then $I_{HXR}$ $\propto$ $V_{CME}$. This is consistent with the result 
of Yashiro \& Gopalswamy (2009). A correlation between the energy release rate, which is represented by $I_{HXR}$, and the 
plasmoid velocity $V_{pl}$ is also successfully reproduced by a magnetohydrodynamic simulation \citep{nisd09}. In the 
simulation, they clearly showed that the plasmoid velocity controls the energy release rate (i.e. reconnection rate) in 
the nonlinear evolution.

Observations of plasmoid ejections have been paid attention to as evidence of magnetic reconnection, though they 
found only one plasmoid ejection per one solar flare. However, magnetic reconnection theory suggests that the impulsive bursty 
regime of reconnection or fractal reconnection is associated with a series of plasmoids of various scales. 
It is known that magnetic reconnection is an effective mechanism for energy release in a solar flare. Once the current sheet 
becomes thin enough for the tearing instability to occur, repetitive formation of magnetic islands and their subsequent 
coalescence drives the ``impulsive bursty'' regime of reconnection \citep{fin77, taj87, pri85}. Furthermore, such reconnection 
can produce a fractal structure in the current sheet, which is not only theoretically predicted but has also been observed 
\citep[i.e.][]{shi01, nis09}. The plasmoids generated in this cycle control the energy release by inhibiting magnetic 
reconnection in the current sheet and/or by inflow driven by the ejection. The plasmoid ejection enhances reconnection and 
promotes further plasmoid ejections from the current sheet. Similar intermittent energy release from a solar flare has been 
reported as multiple downflows associated with hard X-ray bursts in the impulsive phase \citep{asa04a}. McKenzie et al. (2009) 
estimated the total amount of energy of multiple downflows and showed that it is comparable to the total amount of energy 
released from the magnetic field, which is consistent with the magnetic reconnection model. Tanuma et al. (2001) showed 
through numerical simulations that plasmoid ejections are closely coupled with the reconnection process and the greatest 
energy release occurs when the largest plasmoid is ejected. If the strong energy release corresponds to magnetic reconnection, 
we may conclude that this is evidence of unsteady magnetic reconnection in a solar flare and that plasmoids have a key role 
in energy release and particle acceleration.

\begin{figure}
\epsscale{.80}
\plotone{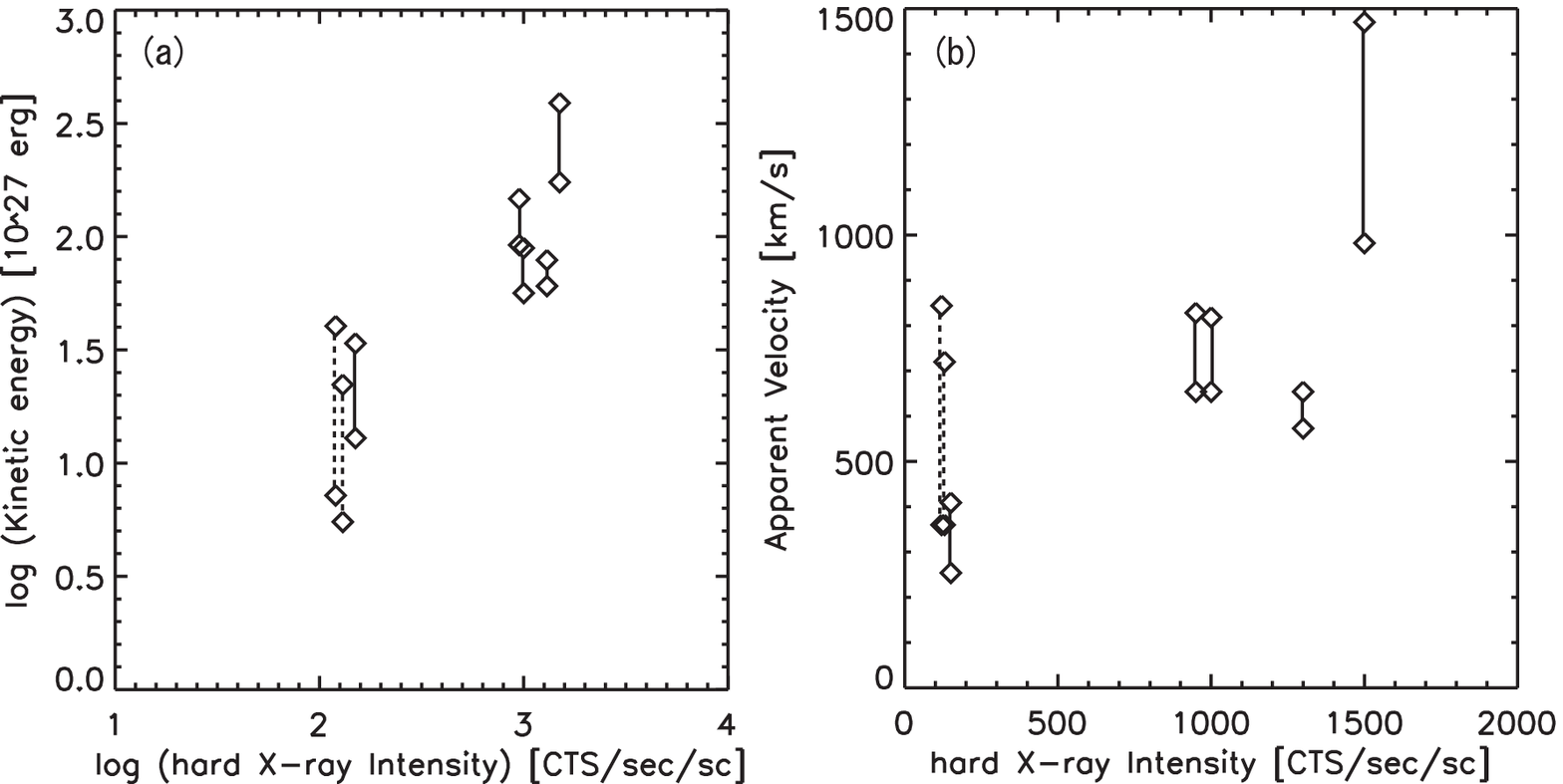}
\caption{(a) Correlation plot between the kinetic energy of plasmoid ejection and the intensity of the corresponding hard 
X-ray peak emission. (b) Correlation plot between the apparent velocity of plasmoid ejection and the intensity of the 
corresponding hard X-ray peak emission.\label{fig10}}
\end{figure}


%
%
%
%

\acknowledgments

We first acknowledge an anonymous referee for his/her useful comments and suggestions. We also thank A. Hillier for his 
careful reading and correction of this paper. We also thank J. Kiyohara for her help in making the movies and T. Morimoto for 
his help in finding the multiple plasmoid ejection events. We made extensive use of YOHKOH/SXT and HXT data. This work was 
supported in part by the Grant-in-Aid for Creative Scientific Research ``The Basic Study of Space Weather Prediction'' 
(Head Investigator: K.Shibata) from the Ministry of Education, Culture, Sports, Science, and Technology of Japan, and in part 
by the Grand-in-Aid for the Global COE program ''The Next Generation of Physics, Spun from Universality and Emergence'' from 
the Ministry of Education, Culture, Sports, Science, and Technology (MEXT) of Japan.\\

\clearpage

\begin{deluxetable}{ccccccccccccrl}
\tabletypesize{\scriptsize}
\rotate
\tablecaption{Physical parameters of the Multiple Plasmoid Ejections\label{tbl-1}}
\tablewidth{0pt}
\tablehead{
\colhead{\#} & \colhead{Time} & \colhead{log T} & \colhead{log EM} & \colhead{Size} &
\colhead{Density} & \colhead{Mass} & \colhead{Velocity} & \colhead{E$_{kin}$} & \colhead{E$_{th}$} & \colhead{Direction} & 
\colhead{I$_{HXR}$}}
\startdata
& (UT) & (K) & (cm$^{-3}$) & (10$^{18}$cm$^2$) & (10$^9$cm$^{-3}$) & (10$^{13}$g) & (km/s) & (10$^{28}$erg) & (10$^{28}$erg) & 
& (CTS/sec/sc) \\
P1 & 15:09:19-15:09:59 & 6.98-7.09 & 45.32-45.33 & $>$3.9-4.7 & 2.76-2.79 & $>$2.63-2.66 & 654-818  & $>$5.63-8.91 & 
$>$2.51-3.28 & NorthWest & 1000 \\
P2 & 15:09:59-15:10:59 & (7.00) & (45.34) & $>$4.6-5.8 & (2.65) & $>$(3.69) & 573-654 & ($>$6.05-7.88) & 
($>$3.69) & NorthWest & 1300 \\ 
P3 & 15:10:19-15:10:59 & (7.00) & (45.32) & $>$4.7-8.3 & (2.59) & $>$(3.60) & 982-1470 & ($>$17.4-38.9) & 
($>$3.60) & NorthWest & 1500 \\
P4 & 15:10:39-15:11:19 & 7.02-7.07 & 45.80-45.80 & $>$3.1-7.1 & 4.85 & $>$4.30 & 654-828 & $>$9.19-14.7 & 
$>$4.50-5.05 & NorthWest & 950 \\
P5 & 15:13:33-15:14:27 & 7.02-7.08 & 45.73-45.74 & $>$1.5-3.1 & 5.76-5.83 & $>$1.12-1.13 & 360-844  & $>$0.72-4.03 & 
$>$1.17-1.36 & NorthWest & 120 \\
P6 & 15:13:33-15:13:51 & 7.07-7.11 & 45.49-45.50 & $>$1.6-5.2 & 4.37-4.42 & $>$0.85-0.86 & 360-720  & $>$0.55-2.22 & 
$>$0.99-1.11 & SouthWest & 130 \\
P7 & 15:16:27-15:18:43 & 6.97-7.03 & 46.05-46.06 & $>$3.2-7.8 & 6.95-7.03 & $>$3.99-4.04 & 254-409  & $>$1.29-3.38 & 
$>$3.72-4.33 & NorthWest & 150 \\
Total & - & - & - & - & - & - & - & 43.5-84.7 & 20.2-22.4 &&&& \\
\tableline
Flare & 15:18:33-15:18:53 &  6.73-7.08 & 48.27-48.96 & 9.33 & 137-304 & 391-865 & - & - & 210-1040 &&&&\\
\enddata
\tablecomments{We assumed that the line-of-sight width of a plasmoid is equal to the square-root of the size. As for plasmoid 
ejections P2 and P3, there are no Be/thick Al filter images so that we assumed temperature of the plasmoids as $10^7$ K. We used 
two images of thick Al filter just before and after the one of Be filter. The error in Table 1 comes from the result of these two 
images. This error is much greater than that of photon noise (see Ohyama \& Shibata 1996 in more detail).
}
\end{deluxetable}

\clearpage



\end{document}